\documentclass[aps,prd,showpacs,preprintnumbers]{revtex4}
\usepackage{graphics}
\def \beq{\begin{equation}}
\def \eeq{\end{equation}}
\def \beqa{\begin{eqnarray}}
\def \eeqa{\end{eqnarray}}
\def \cv{c_{\scriptscriptstyle V}}

\def \th{T_H}

\def \ie{{\sl i.e.\/}}
\def \etal{{\sl et al.\/}}

\def \jp{{\sl J.\ Phys.\/}}

\def \pl{{\sl Phys.\ Lett.\/}}
\def \pr{{\sl Phys.\ Rev.\/}}
\def \prl{{\sl Phys.\ Rev.\ Lett.\/}}
\def \zp{{\sl Z.\ Phys.\/}}

\begin{document}
\title{Stabilizing Hadron Resonance Gas Models against Future Discoveries}
\author{S.\ \surname{Chatterjee}}
\email{sandeep@cts.iisc.ernet.in}
\affiliation{Center for High Energy Physics,\\ Indian Institute of Science,\\
         Bangalore 560012, India.}
\author{R.\ M.\ \surname{Godbole}}
\email{rohini@cts.iisc.ernet.in}
\affiliation{Center for High Energy Physics,\\ Indian Institute of Science,\\
         Bangalore 560012, India.}
\author{Sourendu \surname{Gupta}}
\email{sgupta@tifr.res.in}
\affiliation{Department of Theoretical Physics, Tata Institute of Fundamental
         Research,\\ Homi Bhabha Road, Mumbai 400005, India.}

\begin{abstract}
We examine the stability of hadron resonance gas models by extending them
to take care of undiscovered resonances through the Hagedorn formula. We
find that the influence of unknown resonances on thermodynamics is
large but bounded. Hadron resonance gases may be internally consistent
up to a temperature higher than the cross over temperature in QCD;
but by examining quark number susceptibilities we find that their
region of applicability seems to end substantially below the QCD cross
over. We model the decays of resonances and investigate the ratios of
particle yields in heavy-ion collisions. We find that observables such
as hydrodynamics and hadron yield ratios change little upon extending the
model. As a result, heavy-ion collisions at RHIC and LHC are insensitive
to a possible exponential rise in the hadronic density of states, thus
increasing the stability of the predictions of hadron resonance gas
models in this context.
\end{abstract}
\maketitle

\begin{figure}[t]
\begin{center}
   \scalebox{0.65}{\includegraphics{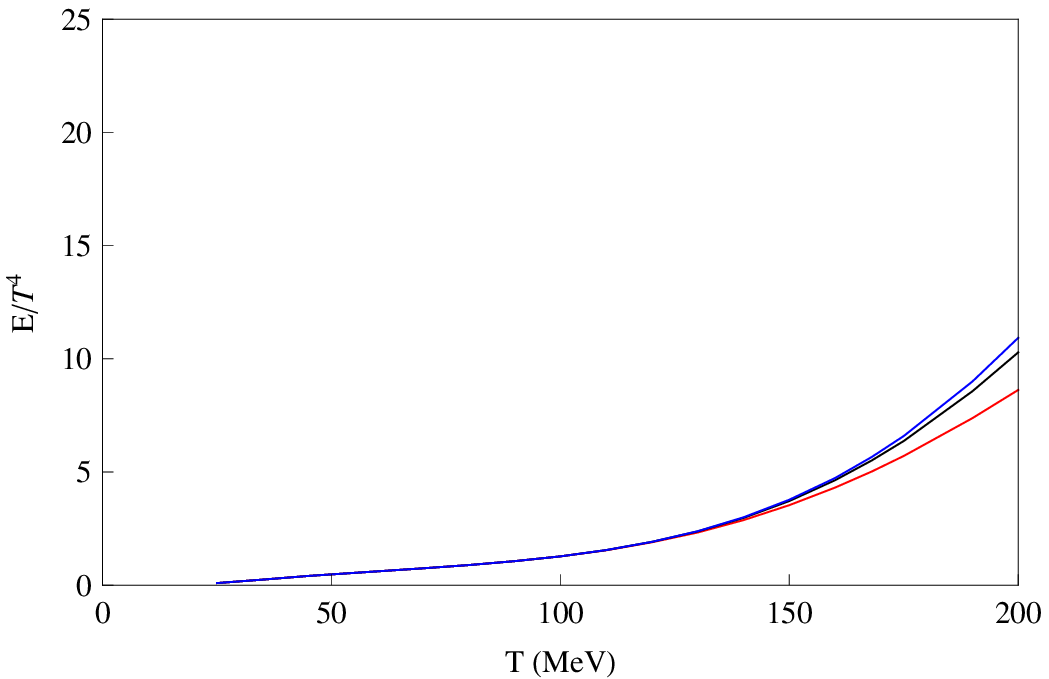}}\hfil
   \scalebox{0.65}{\includegraphics{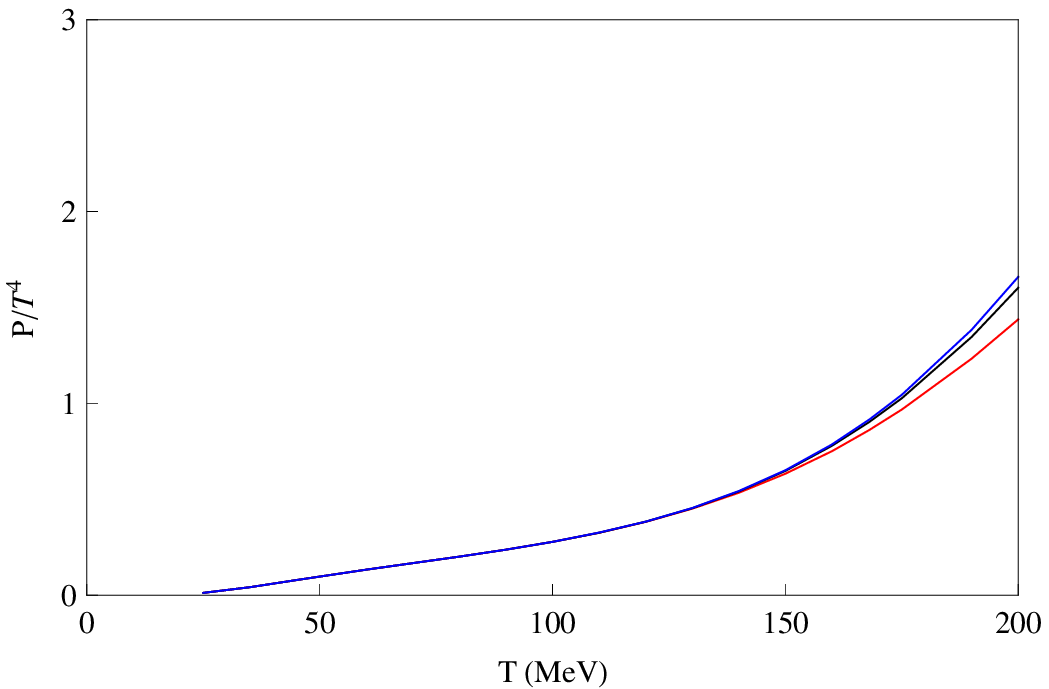}}
   \scalebox{0.65}{\includegraphics{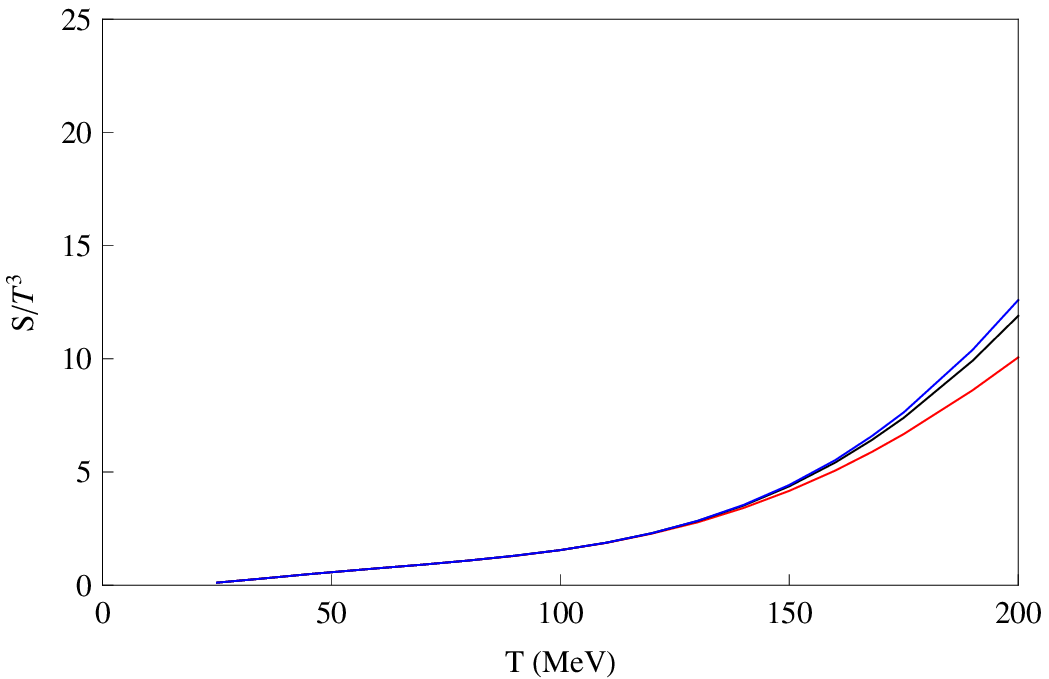}}\hfil
   \scalebox{0.65}{\includegraphics{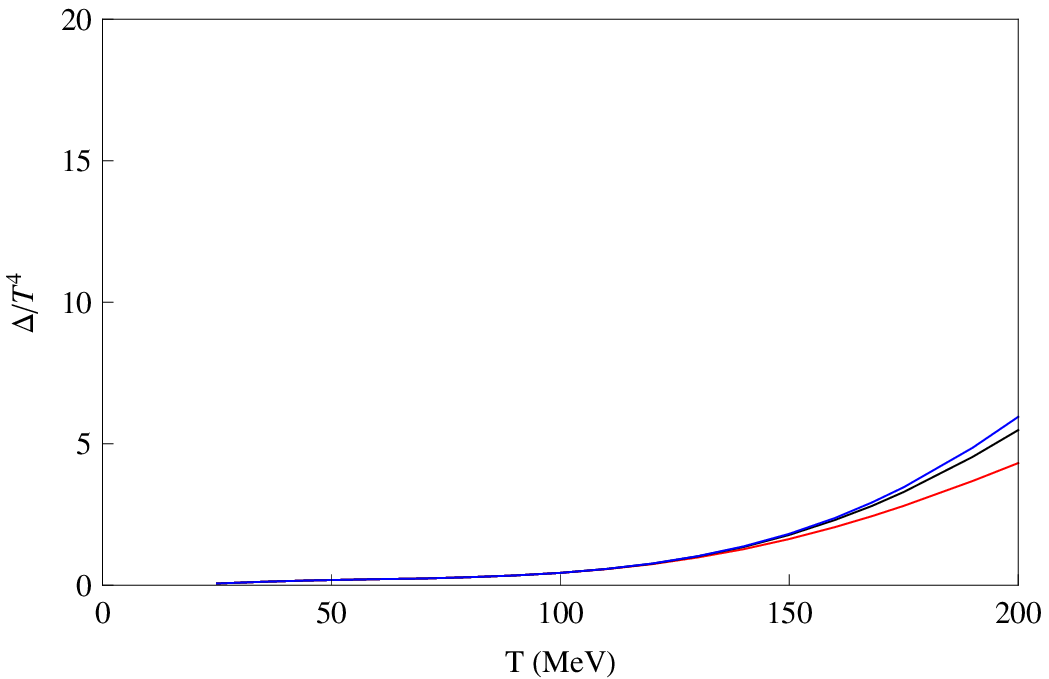}}
   \scalebox{0.65}{\includegraphics{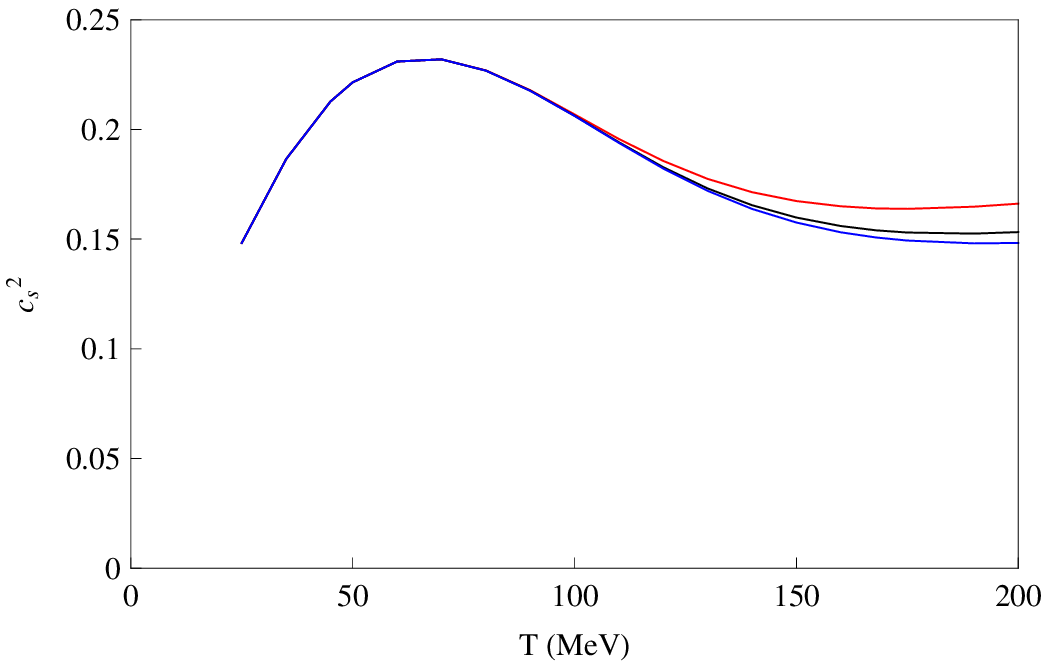}}\hfil
   \scalebox{0.65}{\includegraphics{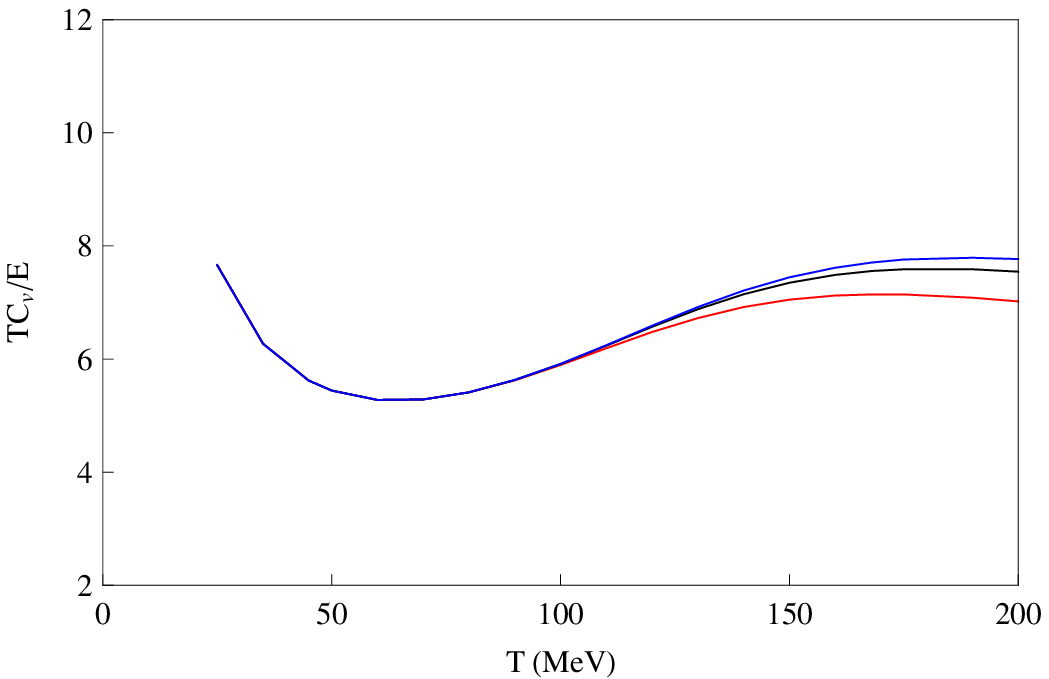}}
\end{center}
\caption{Thermodynamic quantities in the hadron resonance gas variant which
  we call HRG1. As the mass cutoff, $\Lambda$ on the states included in the
  model is changed the predictions change monotonically. These are results
  at zero chemical potential.}
\label{fg.hrg1}
\end{figure}

\begin{figure}
\begin{center}
   \scalebox{0.65}{\includegraphics{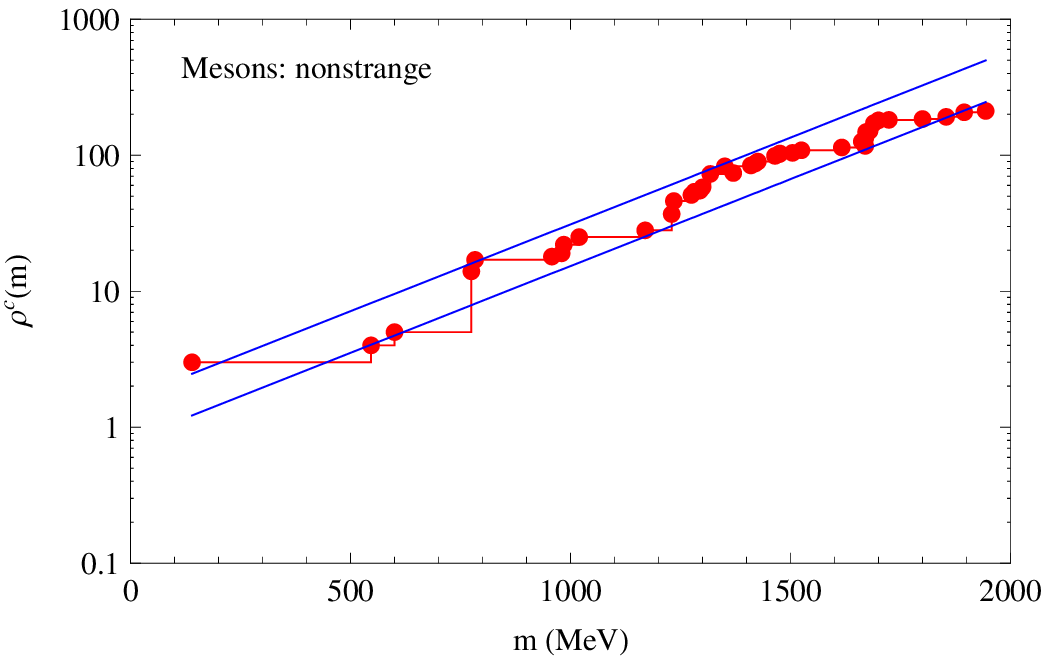}}\hfil
   \scalebox{0.65}{\includegraphics{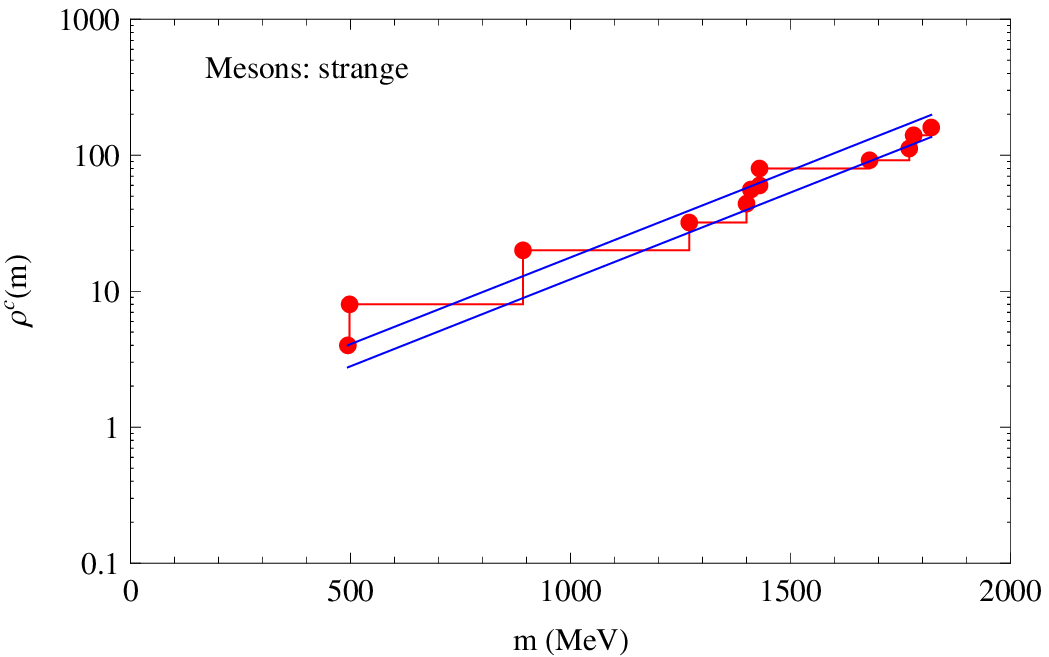}}
   \scalebox{0.65}{\includegraphics{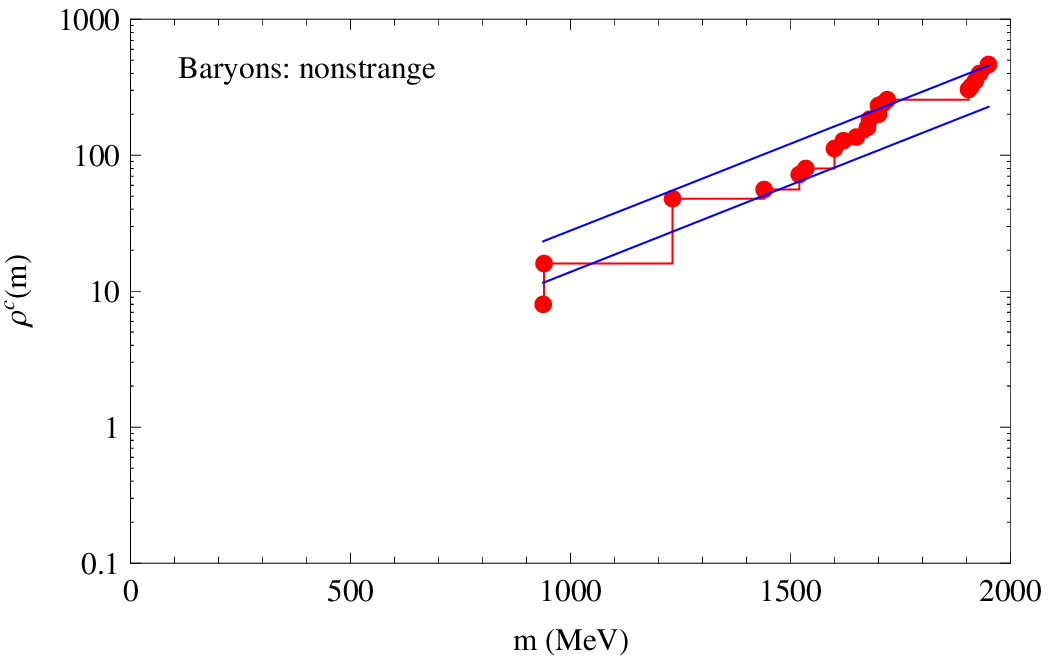}}\hfil
   \scalebox{0.65}{\includegraphics{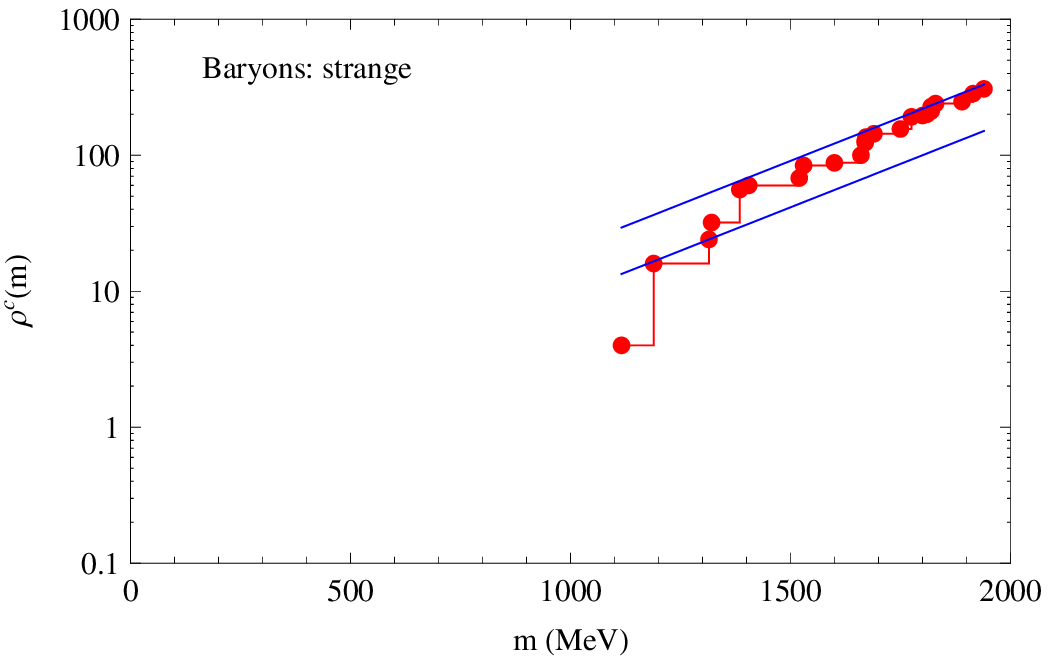}}
\end{center}
\caption{Fit of the parameters of the Hagedorn spectrum to states from the
  particle data book for masses up to 2000 MeV. The band is wide enough to
  include all but the two extreme points in the cumulative distribution
  for each of four classes of hadrons.}
\label{fg.hagfit}
\end{figure}

Hadron resonance gas models (HRGM) in their modern form were first
explored in \cite{dixit,suhonen,greiner,heinz}, where the emphasis was
on exploring the phase diagram.  Then, following the CERN SPS experiments
with heavy-ions, it was found that integrated particle yields could be
explained in HRGMs \cite{bms,yen,becattini}.  Typically, one defines
the thermodynamics of HRGMs through the summed free energy---
\beq
   \log Z(V,T,\mu) = \int dm\left[\rho_M(m) \log Z_b(m,V,T,\mu)
      + \rho_B(m) \log Z_f(m,V,T,\mu)\right],
\label{hrgm}\eeq
where the gas is contained in a volume $V$, has temperature $T$ and
chemical potential $\mu$, $Z_b$ is the partition function for an ideal gas
of bosons with mass $m$, $Z_f$ of an ideal gas of fermions, $\rho_M(m)$ is
the spectral density of mesons, and $\rho_B(m)$ of baryons. From this one
computes the energy density, $E$, by taking a derivative with respect to
$1/T$, and the pressure, $P$, by taking a derivative with respect to $V$.
One can also find the conformal symmetry breaking measure $(E-3P)/T^4$,
the entropy density, $S/T^3=(E+P)/T^4$, the speed of sound, $c_s^2 =
dP/dE$, and the specific heat, $\cv$.

Hadron properties enter these models through $\rho_{B,M}$ and
the treatment of the decay width and size of the hadron. We follow
\cite{becattini} in neglecting an excluded volume effect for hadrons,
\ie, we treat the hadrons as effectively point-like. In constructing
the thermodynamics, the decay width is usually neglected; we accept
this assumption through the part of this work which deals with the
equation of state, but, as usual, include decays when we explore other
aspects of HRGMs. One aspect common to models of the type explored in
\cite{bms,yen,becattini} is to take the observed spectrum of hadrons up
to some cutoff $\Lambda$, \ie, write
\beq
   \rho_{M,B}(m) = \sum_i^{m_i\le\Lambda} g_i\delta(m-m_i),
\label{hg1}\eeq
where $m_i$ are the masses of the known hadrons and $g_i$ the degeneracy
factor $(2J_i+1)(2I_i+1)$ where $J_i$ is the spin and $I_i$ the isospin,
and the sum is over meson or baryon states, as appropriate. We call
such models HRG1. We include in the sum above all the states which are
reasonably well established, and labeled as better than 1 star in the
particle data book.

Clearly the predictions of thermodynamic quantities in HRG1 depend
on the mass cutoff $\Lambda$. This is shown in Figure \ref{fg.hrg1}.
Even at temperatures as low as 150 MeV, there is a 5\% increase in the
energy density when $\Lambda$ is increased from 1600 MeV to 1800 MeV,
and a further 2\% increase when it is increased to 2000 MeV. Changes at
temperatures of 170 MeV or so are significantly larger.  Predictions of
quantities such as the speed of sound, $c_s^2$, and the specific heat,
$\cv$, are significantly less stable. Recent quark model \cite{qmod}
and lattice computations \cite{lat} lead us to believe that there is a
much higher density of hadronic states in the mass range 2--3 GeV than
below 2 GeV. If so, it is not clear by how much the predictions will
change as $\Lambda$ is increased to 3 GeV or beyond.

In order to explore the stability of predictions from HRGMs, we develop
a variant of these models in which we take the observed spectrum of
states up to a certain cutoff $\Lambda$, as before, and above this we
put in an exponentially rising cumulative density of hadron states in
\cite{hagedorn,frautschi}. It is well known that such a density of states
overcomes the exponential suppression of higher mass resonances in eq.\
(\ref{hrgm}). As a result, there is a limiting temperature in
these models, which was, long ago, taken as evidence of a QCD phase
transition \cite{cabibbo}. So, the extended model, HRG2, has
\beq
   \rho_h(m) = \sum_i^{m_i\le\Lambda} g_i\delta(m-m_i)
        + \frac{a_h}{\th} {\rm e}^{m/\th} \Theta(m-\Lambda),
\label{hg2}\eeq
where the model parameters $\th$ and $a_h$ are fitted to data on the
cumulative distribution of different sets of hadrons, $h$.  Similar models
have been used to study observables as diverse as dilepton production
\cite{ruuskanen} to chemical equilibration \cite{shovkovy}.  A comparison
of the equation of state with different density of states for the Hagedorn
spectrum is presented later.

\begin{table}[htb]
\begin{center}
\begin{tabular}{|c|c|c|}
\hline
Class ($h$) & $a_h^{\rm min}$ & $a_h^{\rm max}$ \\
\hline
Mesons (non strange) & 0.81 & 1.64 \\
Mesons (strange) & 0.64 & 0.94 \\
Baryons (non strange) & 0.73 & 1.47 \\
Baryons (strange) & 0.50 & 1.10 \\
\hline
\end{tabular}
\end{center}
\caption{The extreme values of $a_h$ such that for $\th=340$ MeV the
  cumulative density of hadron states up to 2000 MeV lies within the bands
  except for the two extreme points on each side.}
\label{tb.hagfit}
\end{table}

\begin{figure}
\begin{center}
   \scalebox{0.65}{\includegraphics{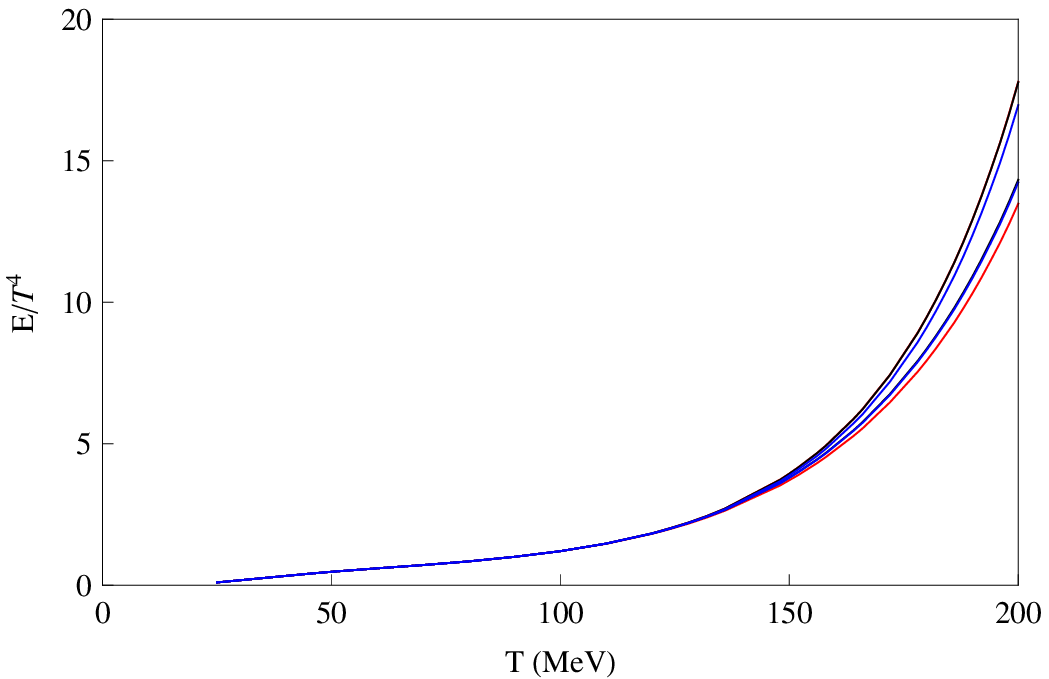}}\hfil
   \scalebox{0.65}{\includegraphics{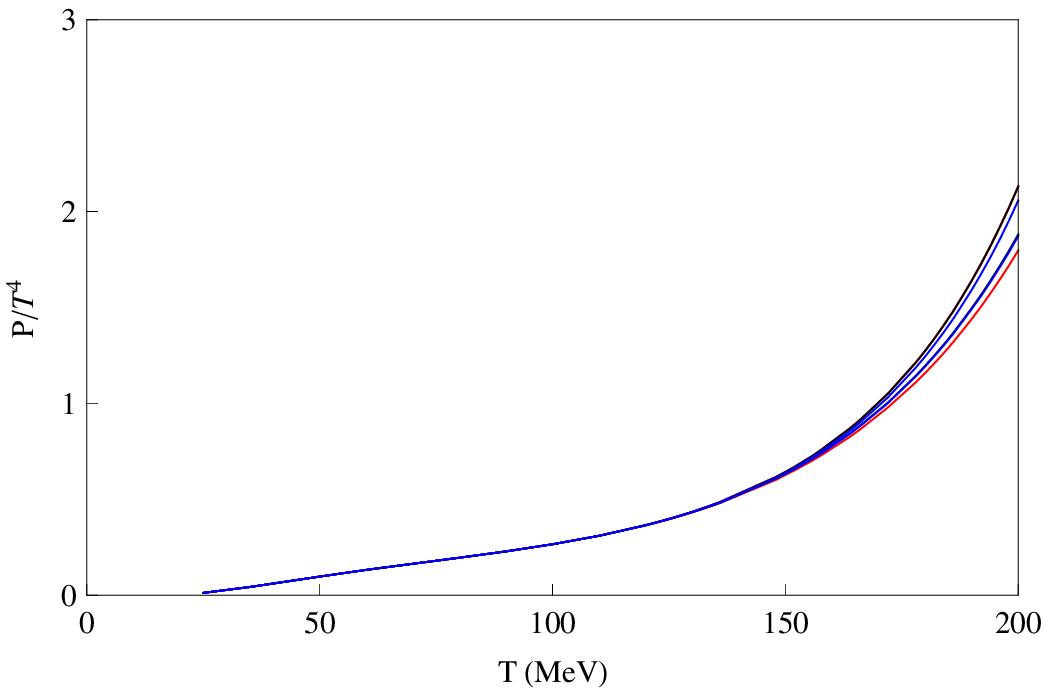}}
   \scalebox{0.65}{\includegraphics{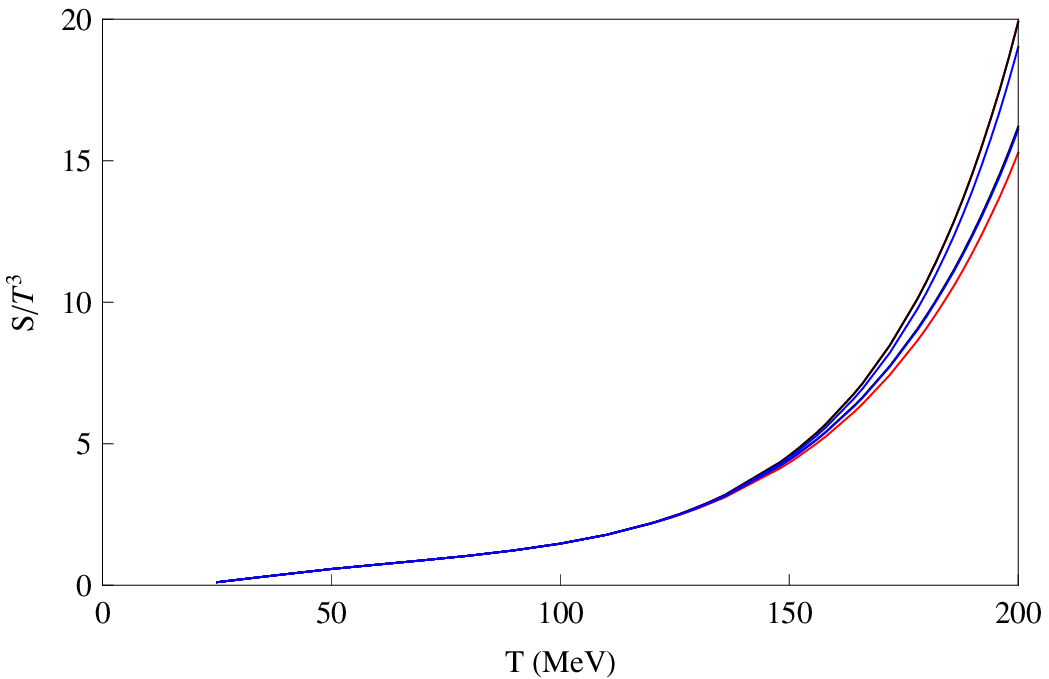}}\hfil
   \scalebox{0.65}{\includegraphics{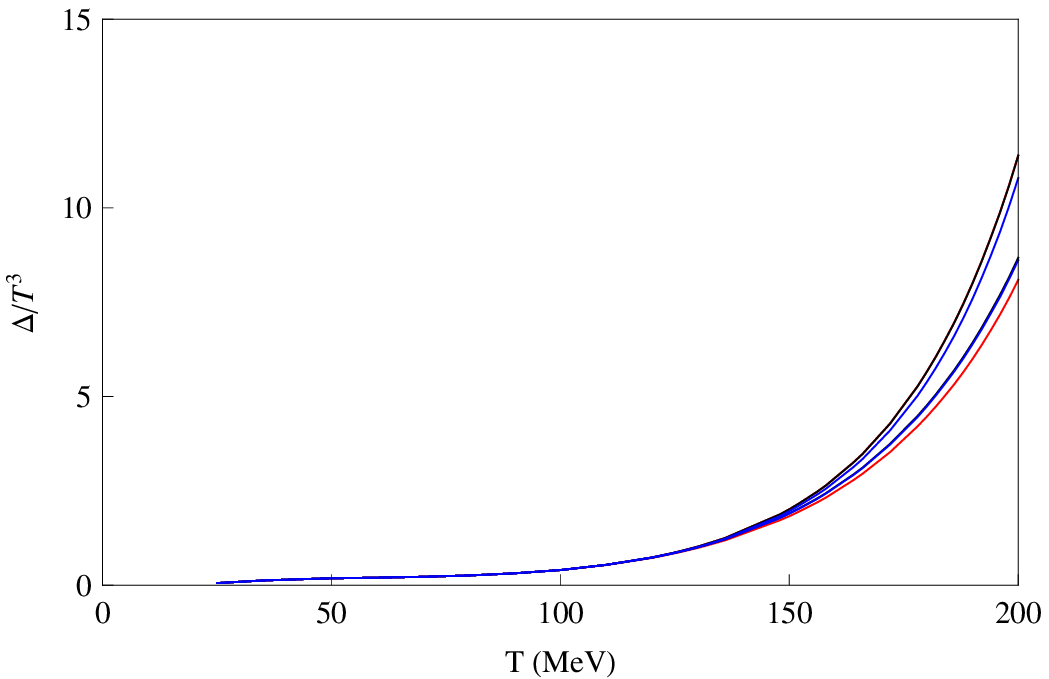}}
   \scalebox{0.65}{\includegraphics{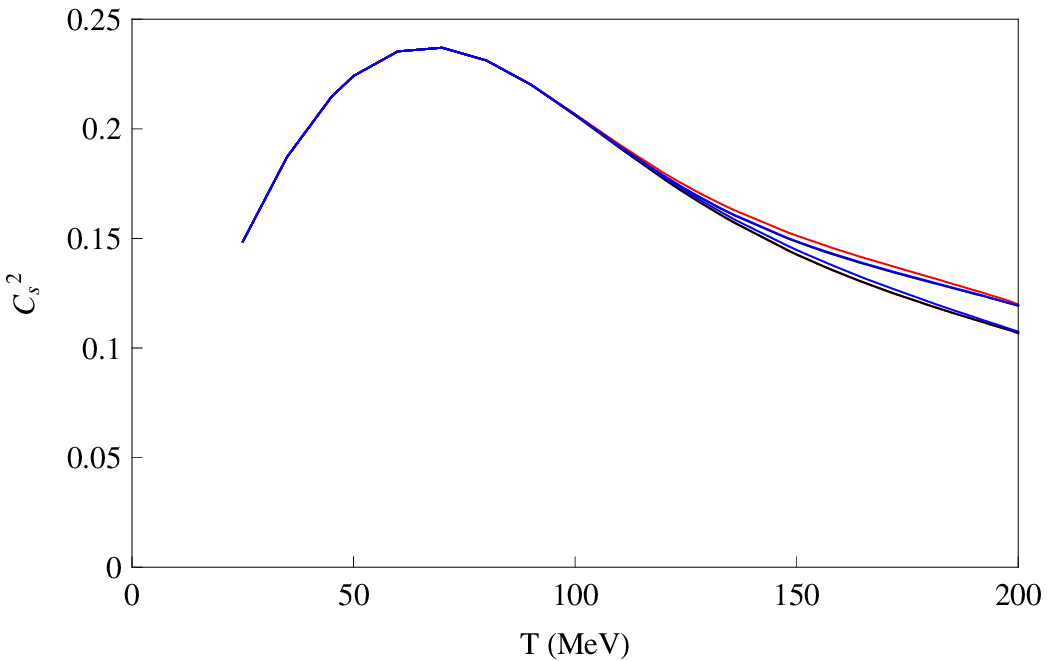}}\hfil
   \scalebox{0.65}{\includegraphics{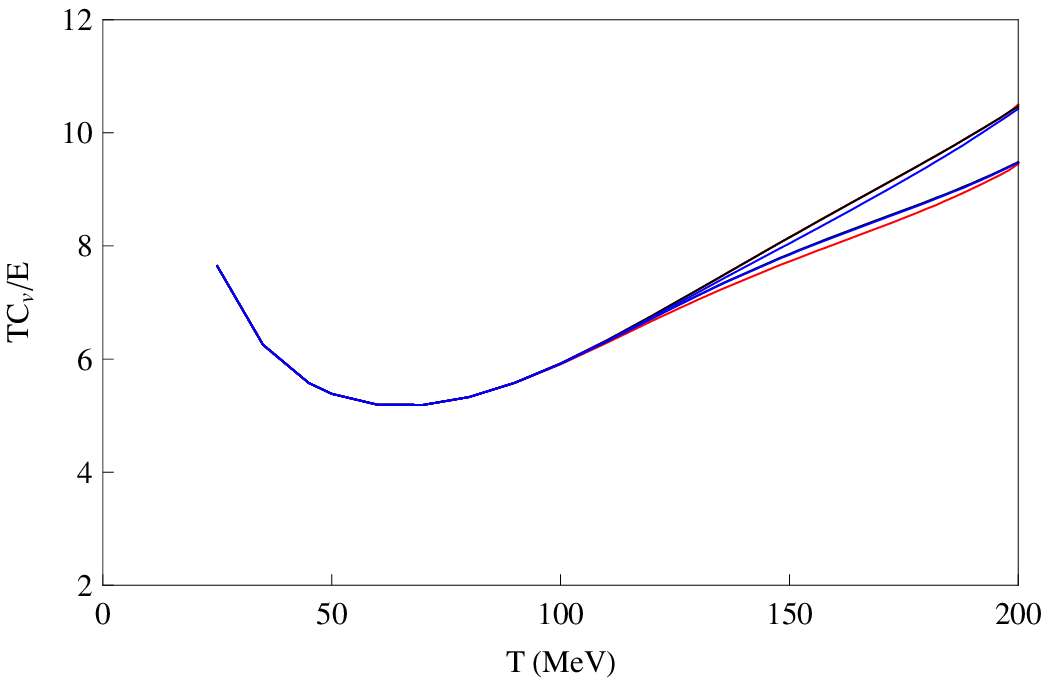}}
\end{center}
\caption{Thermodynamic quantities in the hadron resonance gas variant which
  we call HRG2. At each value of the parameter $\Lambda$, there are upper
  and lower limits on every thermodynamic quantity. The predictions of HRG2
  are stable, since the allowed band shrinks as $\Lambda$ increases, and
  the band for higher $\Lambda$ is entirely contained inside that for a
  lower $\Lambda$. If QCD has a cross over at finite temperature, and the
  same equation of state is valid on both sides, then $c_s^2$ falls
  monotonically across this point, and the softest point is not at $T_c$.}
\label{fg.hrg2}
\end{figure}

\begin{figure}
\begin{center}
   \scalebox{0.65}{\includegraphics{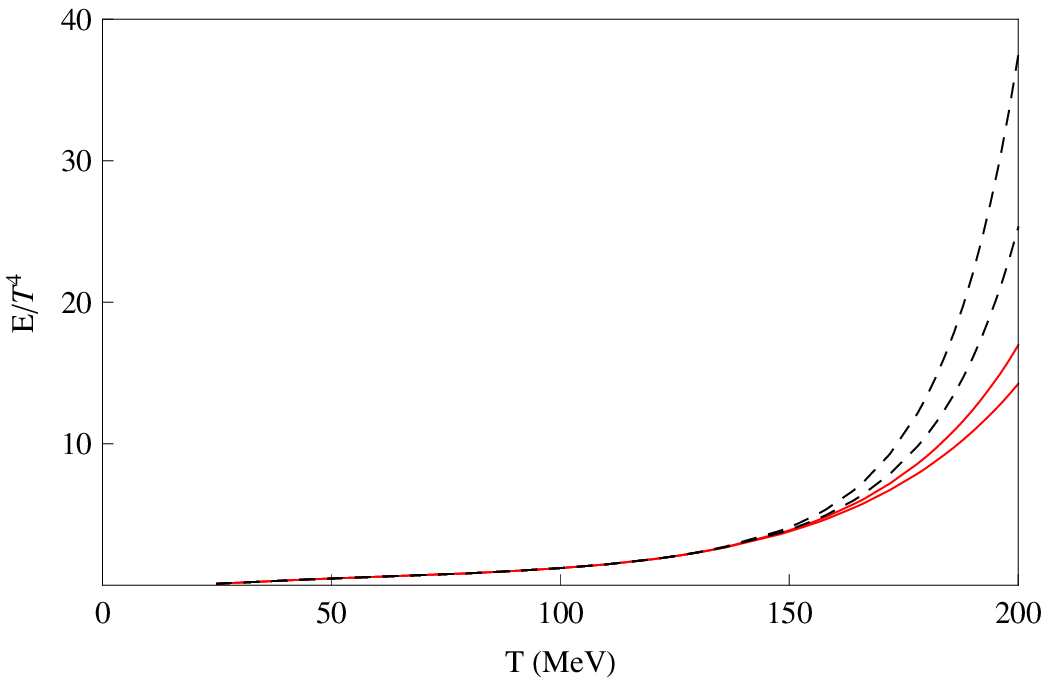}}\hfil
   \scalebox{0.65}{\includegraphics{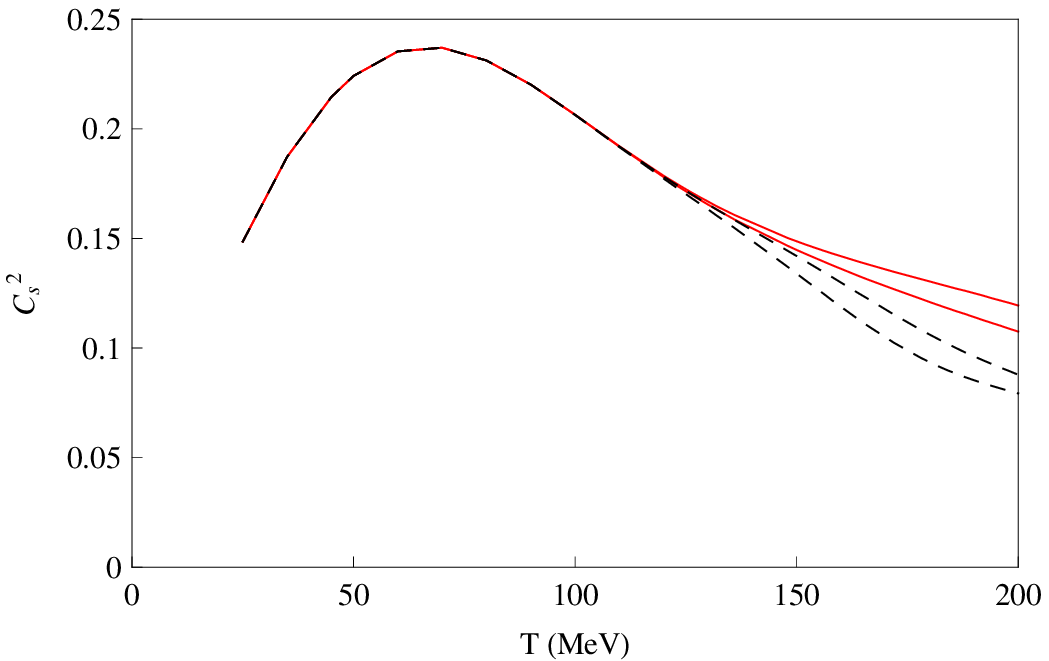}}\hfil
\end{center}
\caption{Comparing predictions of thermodynamics using HRG2 with the
  parametrisation of eq.\ (\ref{hg2}) (full line) and eq.\ (\ref{hg2p})
  (dashed line). The uncertainty bands of all quantities overlap over
  the range of temperatures of interest.}
\label{fg.comparespec}
\end{figure}

We perform fits separately to non-strange mesons, strange mesons,
non-strange baryons and strange baryons. All states in the particle
data book with rating higher than 1 star and mass up to 2000 MeV are
used in these fits.  The quantity $\th$ is obtained by a global fit to
all four types of hadrons, whereas we fit upper and lower values for
$a_h$ for each class of hadrons. We have allowed the band of $a_h/\th$
to be wide enough that all but two of the extreme high and low points
are within the band.  The goodness of such fits can be gauged from Figure
\ref{fg.hagfit}.  The fitted parameter sets to be used in eq.\ (\ref{hg2})
are given in Table \ref{tb.hagfit}. Note that the best fit value of $\th$
is significantly higher than $T_c$, the cross over temperature of QCD
\cite{jaipur}. This is not unexpected, since there is good evidence that
$\th$ and $T_c$ are not equal, at least at finite $N_c$ \cite{bringolz}.
As a result, the model in eq.\ (\ref{hg2}) is internally consistent well
beyond $T_c$.

Using this extended model, we get the predictions of thermodynamic
quantities shown in Figure \ref{fg.hrg2}. Note that for each value of
$\Lambda$ there is an upper and lower limit to the prediction for any
thermodynamic quantity, one from the maximum and the other from the
minimum of the $a_h$.  As one changes $\Lambda$, the band for higher
$\Lambda$ lies entirely within that for lower $\Lambda$. This pleasant
property implies a stability of the predictions of the HRG2 model for
bulk thermodynamic quantities. Even for $c_s^2$ and $\cv$, predictions
of HRG2 are stable in this sense up to a temperature of 160 MeV. The
allowed band at 150 MeV is about 2--7\% for $\Lambda=2000$ MeV.

A different density of states for the Hagedorn model has also been used
in the literature \cite{ranft,polish}. The density of states in the HRG2
with this change would be
\beq
   \rho_h(m) = \sum_i^{m_i\le\Lambda} g_i\delta(m-m_i)
        + \frac{c_h}{(m^2+m_0^2)^{5/4}} {\rm e}^{m/\th} \Theta(m-\Lambda).
\label{hg2p}\eeq
Typically this model is used with $m_0=500$ MeV. We have used this
canonical value as well $m_0=250$ MeV and 1000 MeV. Note that in the limit
as $m_=0\to\infty$, this model reduces to the one in eq.\ (\ref{hg2}). The
best fit value of $\th$ changes significantly as we vary $m_0$, being 210
MeV for the central value of $m_0$, and sliding up to 250 MeV as $m_0$
changes to 1000 MeV. The quality of fit improves marginally as $m_0$
increases, being best for the density of states in eq.\ (\ref{hg2}).
The uncertainty in the value of the derived hadronic quantity, $\th$,
is closely related to the fact that a string model of hadrons is not
a unique and self-consistent theory. As a result, the pre-exponential
factor can be tweaked at will, resulting in large possible changes to
the string tension, or, equivalently, to $\th$.

Using $m_0=500$ MeV and $\th=210$ MeV, we take as the allowed band of
$c_h$ a definition analogous to that used in Table \ref{tb.hagfit}. This
gives us the uncertainties in the corresponding predictions of
thermodynamic quantities. As one could guess, the pleasantly stable
results for thermodynamics shown in Figure \ref{fg.hrg2} are model
artifacts.  In Figure \ref{fg.comparespec} we show that the uncertainty
bands in the two models defined by eqs.\ (\ref{hg2}, \ref{hg2p})
exclude each other in the vicinity of $T=170$ MeV. In view of these
uncertainties, we do not consider the more detailed models which have
been used \cite{polish}. In future, when observations of the hadron
spectrum are extended to significantly larger masses, a more detailed
consideration of string models may become a fruitful topic of research.

Lattice computations of the equation of state give $E/T^4=1.4\pm0.2$ at a
temperature of 140 MeV \cite{jaipur}. At such low temperatures the known
hadron spectrum dominates the results (see Figure \ref{fg.comparespec}).
The mismatch between lattice and hadron gas models is due to the fact
that current day lattice thermodynamics computations are performed on
lattices which are too coarse at such low temperatures. The best lattices
have cutoff of around 1100 MeV when doing thermodynamics at $T=140$ MeV.
Since the cutoff is comparable to the low-lying baryon masses, the hadron
resonance spectrum is strongly disturbed and the low-temperature results
are not yet physical\footnote{The wonderful agreement of a glueball
gas model with lattice computations in pure gauge theory \cite{meyer}
defies this expectation.}. Since the lattice spacing varies inversely
with the temperature, lattice results at higher temperatures are
expected to be reasonable. The efficacy of the lattice closer to $T_c$
is borne out by the fact that renormalization group invariant estimates
of $T_c$ are possible with the cutoffs in use today \cite{scaling}.
Interestingly, present day lattice computations at a temperature of
210 MeV give $E/T^4=12.4\pm0.3$ \cite{jaipur}, which is also below the
prediction of HRG2. However, as shown in Figure \ref{fg.comparespec}, at
this temperature the problem very likely lies with the hadron gas models.
We demonstrate this next.

\begin{figure}
\begin{center}
\scalebox{0.65}{\includegraphics{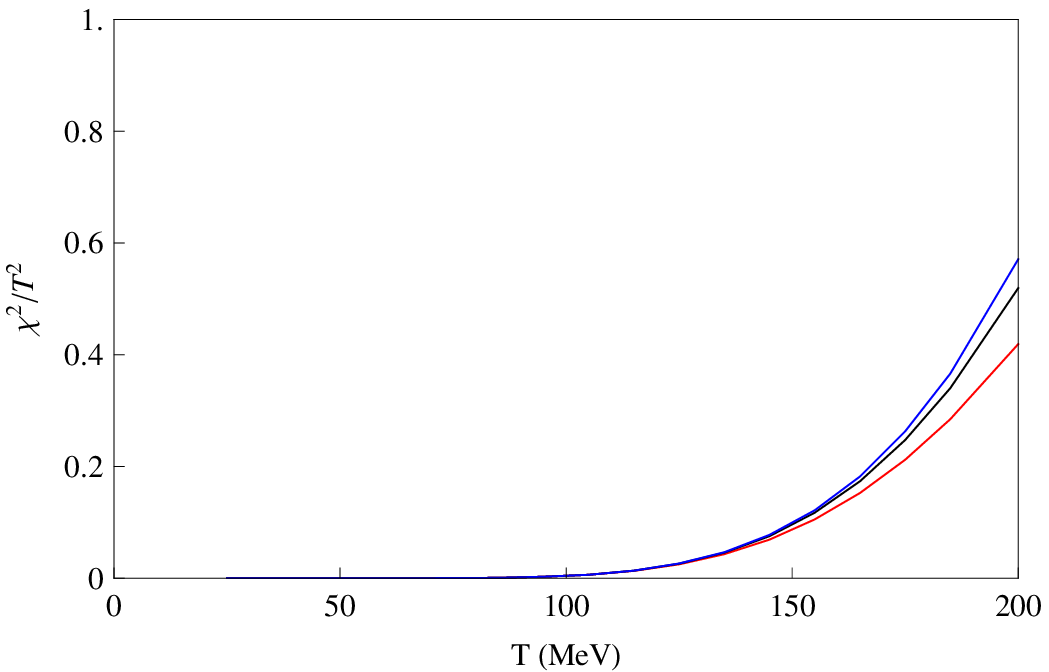}}
\scalebox{0.65}{\includegraphics{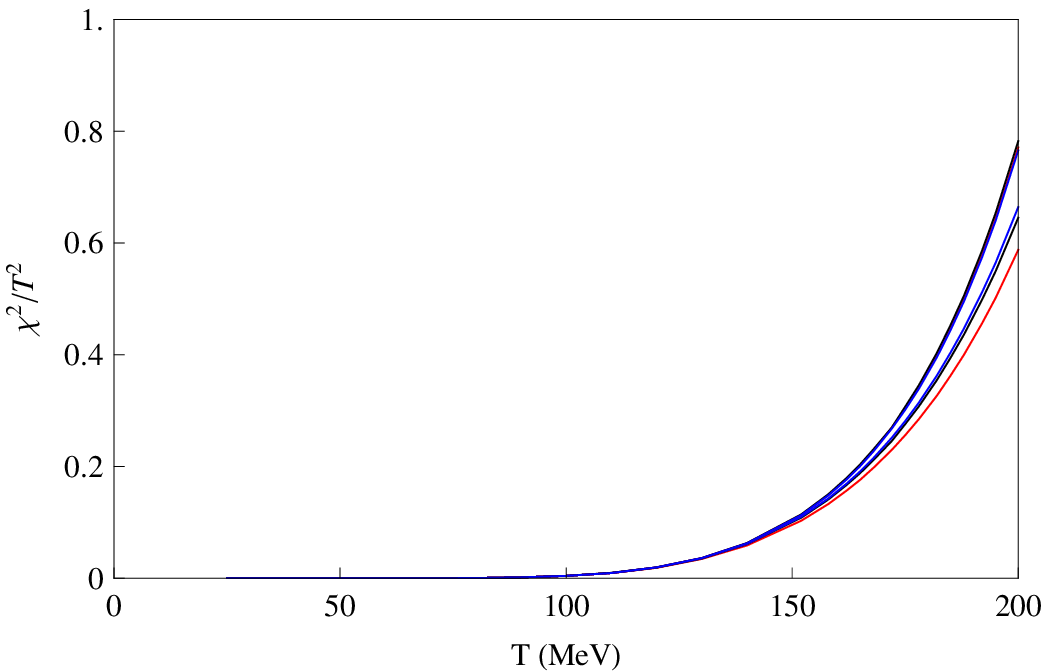}}
\end{center}
\caption{The baryon number susceptibility computed in hadron resonance
  gases. The first panel gives the results in HRG1 and the second in
  HRG2. The colour coding is the same as in previous figures. Note the
  importance of including the Hagedorn spectrum of baryons.}
\label{fg.chi2}
\end{figure}

A phase transition involves a singularity in the free energy. As a
result one usually does not have the same elementary excitations in
terms of which the thermodynamics is constructed on both sides of
a phase transition.  However, there is no singularity of the free
energy at a cross over. As a result, there is no theoretical bar to
a model which is valid on both sides of the cross over.  One way to
investigate the applicability of such models on both sides of $T_c$
is to look at quantum number susceptibilities (QNS). Introduce chemical
potentials for the three quantities conserved in strong interactions,
namely the baryon number, the third component of the isospin ($I_3$) and
the strangeness. Then the baryon number susceptibility is defined to be
\beq
   \chi_B(T) = \left.
      \frac{\partial^2P(T,\mu_B,\mu_{I_3},\mu_S)}{\partial\mu_B^2}
     \right|_{\mu_B=\mu_{I_3}=\mu_S=0}.
\label{qns}\eeq
This defines the susceptibility at zero chemical potential, since there
is lattice data on this quantity, although, of course, it may also be taken
at finite chemical potential. Following \cite{nls} the non-linear
susceptibilities (NLS) are defined as
\beq
   \chi_B^{(n)}(T) = \left.
      \frac{\partial^nP(T,\mu_B,\mu_{I_3},\mu_S)}{\partial\mu_B^n}
     \right|_{\mu_B=\mu_{I_3}=\mu_S=0}.
\label{nls}\eeq
The QNS is just the NLS for $n=2$. In exact analogy one can also define
strangeness susceptibilities and so-called off-diagonal susceptibilities
where some of the derivatives are with respect to one chemical potential
and others with respect to another \cite{strange}.  The QNS predicted by
HRG1 and HRG2 are shown in Figure \ref{fg.chi2}.  The enhanced density of
baryon states in HRG2 leads to a substantial increase in $\chi_B(T)$. As
expected, current lattice data \cite{gg,bi} does not match these curves.

\begin{figure}
\begin{center}
\scalebox{0.65}{\includegraphics{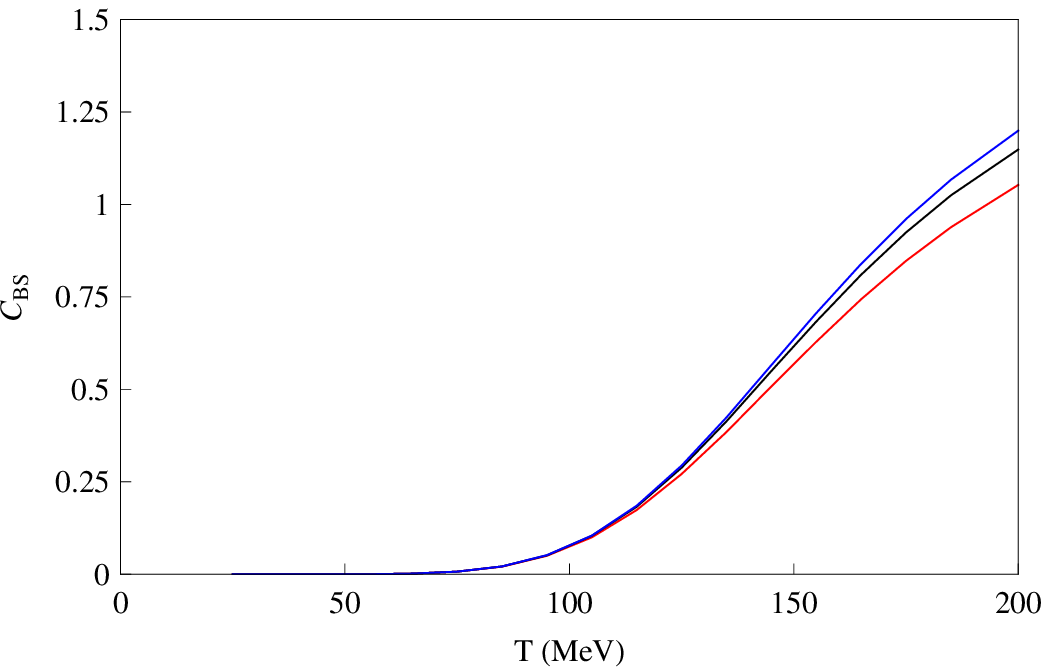}}
\scalebox{0.65}{\includegraphics{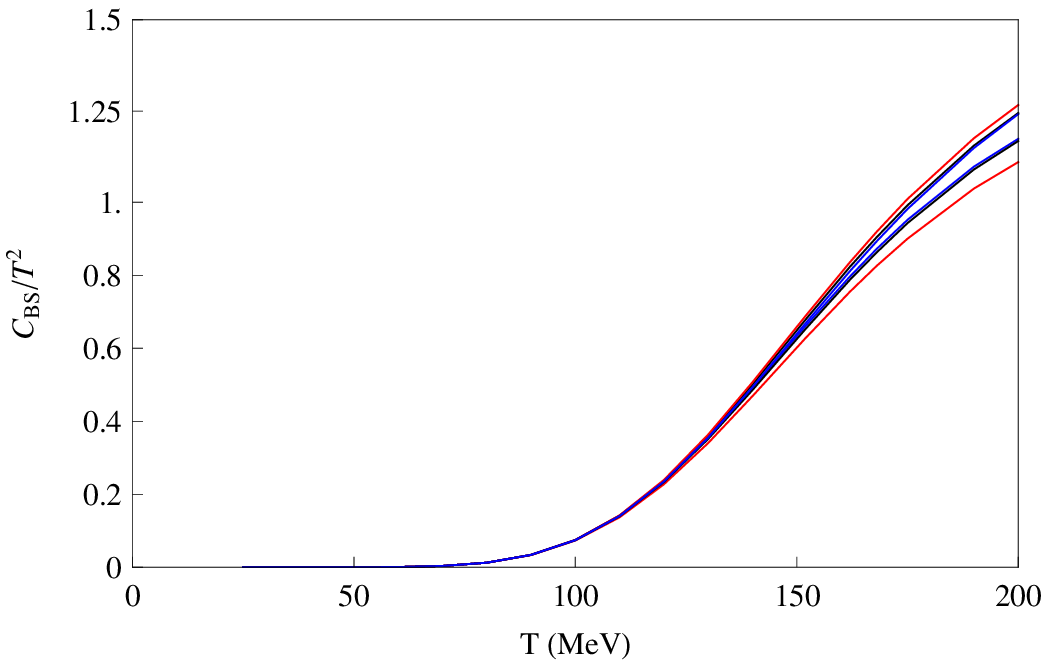}}
\end{center}
\caption{$C_{BS}$ in the hadron resonance gas models HRG1 (left) and HRG2
  (right). One expects $C_{BS}$ to be a monotonic function of temperature
  in a resonance gas model.}
\label{fg.cbs}
\end{figure}

\begin{figure}
\begin{center}
\scalebox{0.40}{\includegraphics{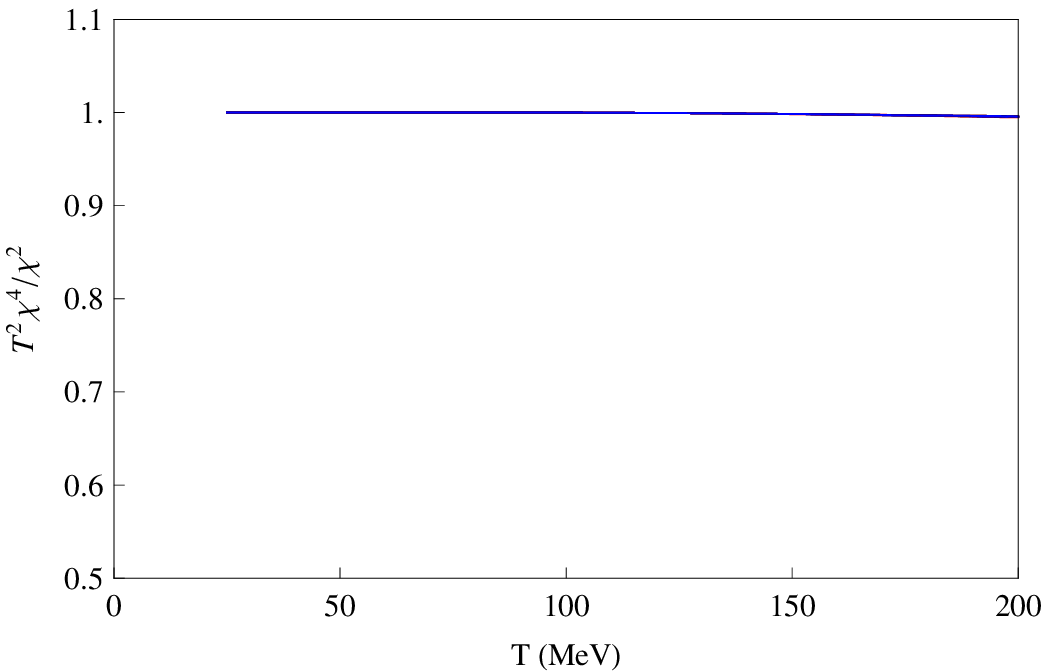}}
\scalebox{0.40}{\includegraphics{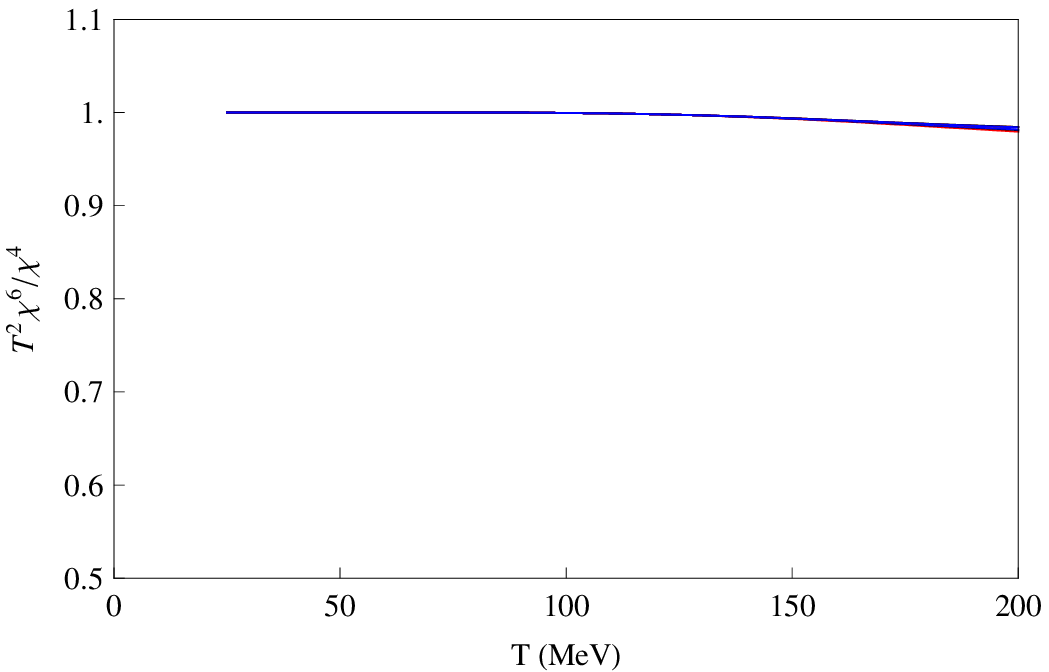}}
\scalebox{0.40}{\includegraphics{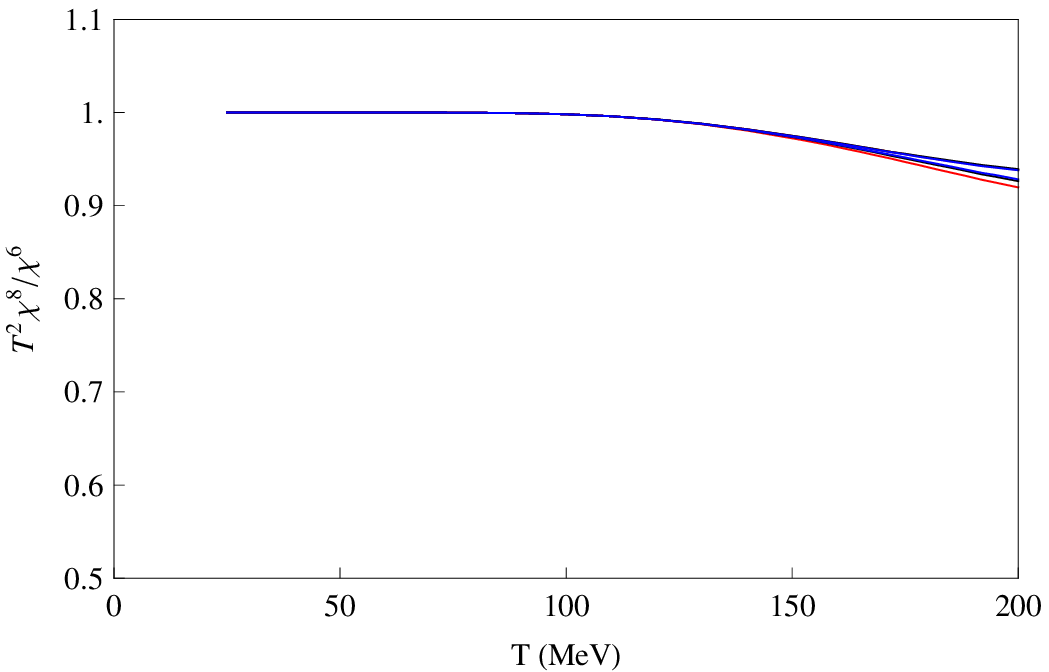}}
\scalebox{0.40}{\includegraphics{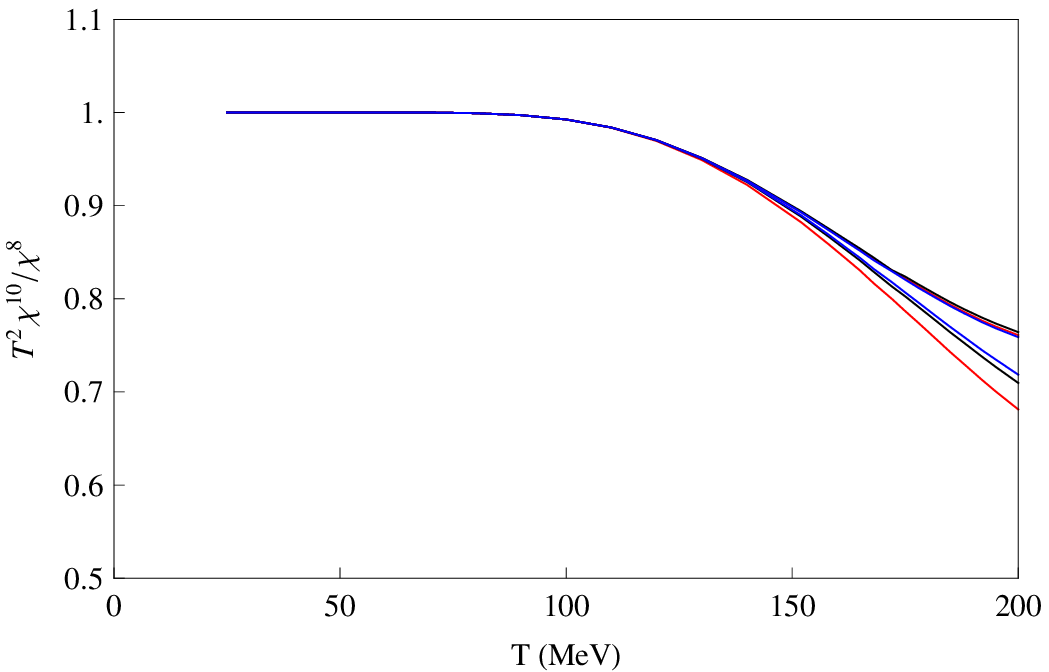}}
\end{center}
\caption{The non-linear baryon number susceptibilities computed in hadron
  resonance gas model HRG2. We have plotted the ratios $\chi^{(n+2)} T^2/
  \chi^{(n)}$ for $n=2$, 4, 6 and 8. The colour coding is the same as
  in previous figures. The results are very stable against the mass
  cutoff.}
\label{fg.chins2}
\end{figure}

The reason is very accurately probed by the quantity
\beq
   C_{BS} = -3 \frac{\chi_{BS}(T)}{\chi_S(T)},
\label{cbs}\eeq
where $\chi_{BS}(T)$ is one of the off-diagonal QNS
\cite{koch,linkage}. The normalization is such that $C_{BS}=1$ in an
ideal gas of quarks.  The HRG1 and HRG2 predictions of $C_{BS}$ are shown
in Figure \ref{fg.cbs}. Since the lowest mass baryons are non-strange,
$C_{BS}$ must start at zero. It is seen to climb monotonically with
temperature. $C_{BS}$ can exceed unity if the contribution of the singly
and doubly strange baryons to $\chi_B$ exceeds half the contribution of
strange mesons to $\chi_S$. Since the strange meson spectrum starts at
a much lower mass than the strange baryon sector, this cannot happen at
low temperature. However, as seen in Figure \ref{fg.hagfit}, the observed
density of strange baryons grows much more rapidly with temperature than
that of strange mesons. As a result, at sufficiently high temperature
$C_{BS}$ exceeds unity.  In HRG2 $C_{BS}$ remains above unity until
$T_H$. Hence there is no continuity between this model and the physics of
the high temperature phase of QCD. 

The ratios of successive NLS in the HRG2 is shown in Figure
\ref{fg.chins2}.  These ratios are not very sensitive to the change
from HRG1 to HRG2. Note that they are extremely constant as a function
of temperature. In contrast, lattice computations show much structure
in the vicinity of $T_c$ as a consequence of the nearness of the QCD
critical point \cite{gg}. The hadron resonance gas, being a mixture
of ideal gases, sees no critical point, but only the Hagedorn limiting
temperature. As a result, it misses all this structure.

We have shown here that there are major points of mismatch between lattice
computations of QNS and the resonance gas models for $T>T_c$.  Since there
is no phase transition in QCD at $\mu_B=0$ but only a crossover, the
failure of a model soon above $T_c$ also implies its failure a little
before $T_c$.  This means that the resonance gas models are restricted
in their range of applicability to well below $T_c$.  In this real QCD
seems to be very different from either pure gauge theory or large-$N_c$
QCD, where the resonance gas model could remain perfectly accurate right
up to the first order phase transition in these models.  We turn next
to the question of whether heavy-ion observables can give information
on the Hagedorn spectrum of resonances.

\begin{figure}
\begin{center}
   \scalebox{0.70}{\includegraphics{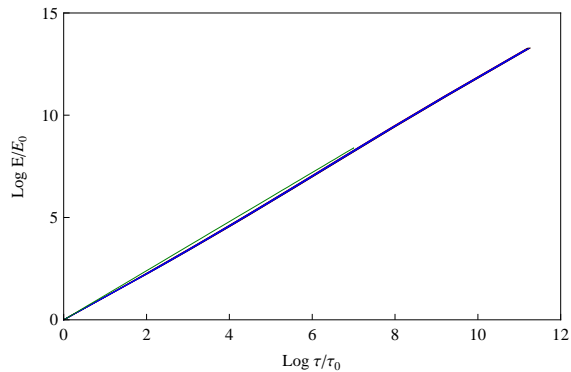}}
\end{center}
\caption{Bjorken flow in the hadron resonance gas phase starting from 
  an initial energy density of 1 GeV/fm$^3$, with the bands of uncertainty
  corresponding to $\Lambda=1600$ MeV, 1800 MeV and 2000 MeV for $c_s^2$.
  Also shown (in green) is the flow for constant $c_s^2=1/5$. There is
  very little difference between these flows.}
\label{fg.bjorken}
\end{figure}

One of the uses to which equations of state can be put is hydrodynamics.
We perform the following simple computation with the HRG2 equation of
state--- use it to evolve a longitudinally expanding fireball which has
cooled to a point where its energy density, $E$, is 1 GeV/fm$^3$. For
longitudinal flow, the quantity $E/E_0$ is a function only of the ratio
$\tau/\tau_0$ where $\tau$ is the proper time, and $\tau_0$ the initial
time, when $E(\tau_0)=E_0$. We integrated the longitudinal flow equation
with upper and lower band of $c_s^2$ for $\Lambda=1600$ MeV, 1800 MeV
and 2000 MeV. We found that the effect of these changes is minimal, as
we show in Figure \ref{fg.bjorken}. For comparison we also show the flow
obtained with constant $c_s^2=1/5$. This is almost indistinguishable
from the other flows. Similarly, the result of a computation with the
density of states in eq.\ (\ref{hg2p}) is indistinguishable from these.
So, hydrodynamics is almost blind to the level of detail in the equation
of state that we have studied.

The main phenomenological application of hadron gas models, however,
is in the analysis of particle yields in heavy-ion collisions. Most
of the particle species observed in the detector are those which are
stable under strong interactions, because the others decay long before
reaching the detectors. The computation of the yield of particles in a
hadron gas model is matched to data to extract the reaction volume and
freezeout temperature and chemical potentials \cite{bms,yen,becattini}.

In HRG1 one creates a table of decays from the particle data book. Each
hadron $H_i$ has decay modes labeled by $\alpha$.  The reaction
\beq
   H_i\to \sum_j n^\alpha_{ij} H_j
\label{reaction}\eeq
proceeds with a branching fraction $B_i^\alpha$, where $n_{ij}^\alpha$
is the number of hadrons $H_j$ produced in this reaction. For later
convenience we add the trivial rule that stable particles decay to
themselves with branching ratio unity. The expected number of $H_j$
produced per decay of $H_i$ are
\beq
   N_{ji} = \sum_\alpha B_i^\alpha n_{ij}^\alpha.
\label{avgdecay}\eeq
For a stable particle $H_i$, we have $N_{ii}=1$ and $N_{ji}=0$ for all
other $j$. In general, some of the decay products, $H_j$, may be unstable
under strong interactions; in that case one has to follow the decay chain
until only stable particles remain. The expected number of $H_j$ resulting
from $H_i$ after all this, ${\cal N}_{ji}$, is easily found by sufficient
number of matrix multiplications---
\beq
   {\cal N}_{ji} = \sum_{k_1,k_2,\cdots,k_m} N_{jk_1} N_{k_1,k_2}
     \cdots N_{k_m,i}, \quad i.e.\quad
   {\cal N} = N^m.
\label{stables}\eeq
The minimum power $m$ to which the matrix $N$ has to be raised is equal to
the maximum number of steps in any decay chain. From the formula above it
is clear that the rows of the matrix $\cal N$ which correspond to unstable
hadrons are zero.  This means that ${\cal N}={\cal N}N$, and hence if $m$
is chosen to be larger than actually required, the cost is in CPU time and
not in correctness. Proofs of all these assertions can be written down
most simply by noting the structure of the matrix $N$, but insight is also
gained by examining a one-to-one correspondence between this problem
and problems on directed graphs \cite{graph}. Finally,
the expected yield of a stable particle, $H_j$, resulting from a fireball
which freezes out at temperature $T$ and chemical potentials $\mu$ is
\beq
   N_j(V,T,\mu) = \sum_i^{m_i<\Lambda} {\cal N}_{ji}
          N^{\rm th}_i(V,T,{\mu}),
\label{yield}\eeq
where $N_i^{\rm th}$ is the number of $H_i$ in thermal equilibrium. From
this, it is clear that for hadron yield computations, it not necessary
to keep track of the detailed decays of every $H_i$. It is sufficient to
keep ${\cal N}_{ij}$, \ie, the expected number of hadrons $H_j$, stable
under strong interactions, resulting from the decay of of each $H_i$.
In any case, the data may be read off the particle data book.

\begin{figure}
\begin{center}
\scalebox{0.65}{\includegraphics{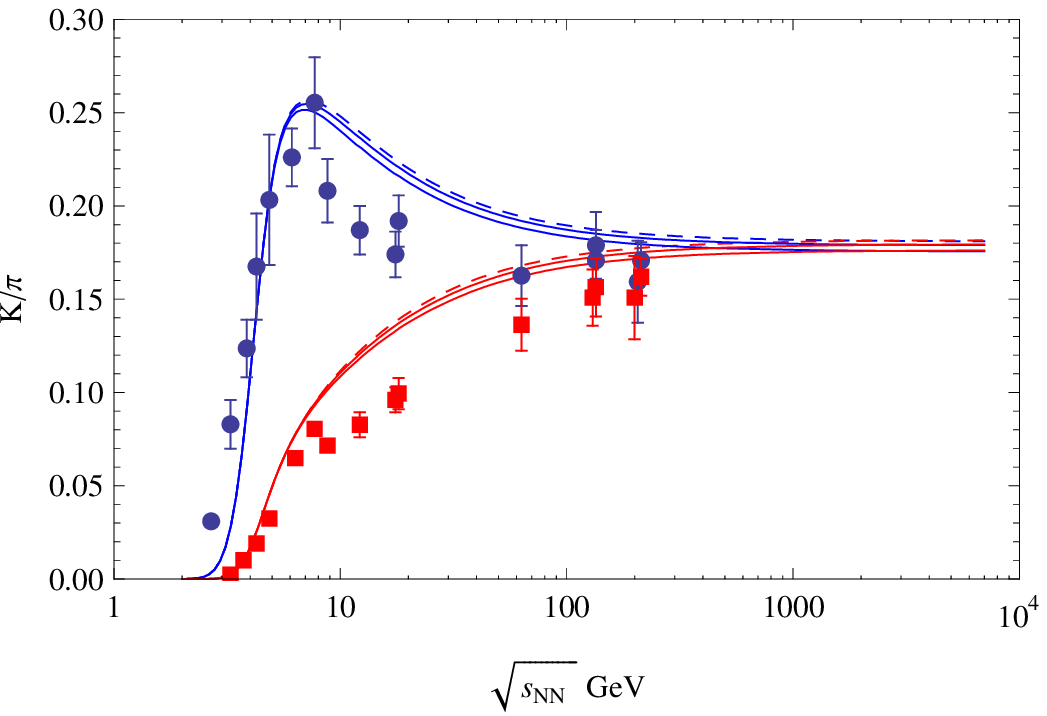}}
\scalebox{0.65}{\includegraphics{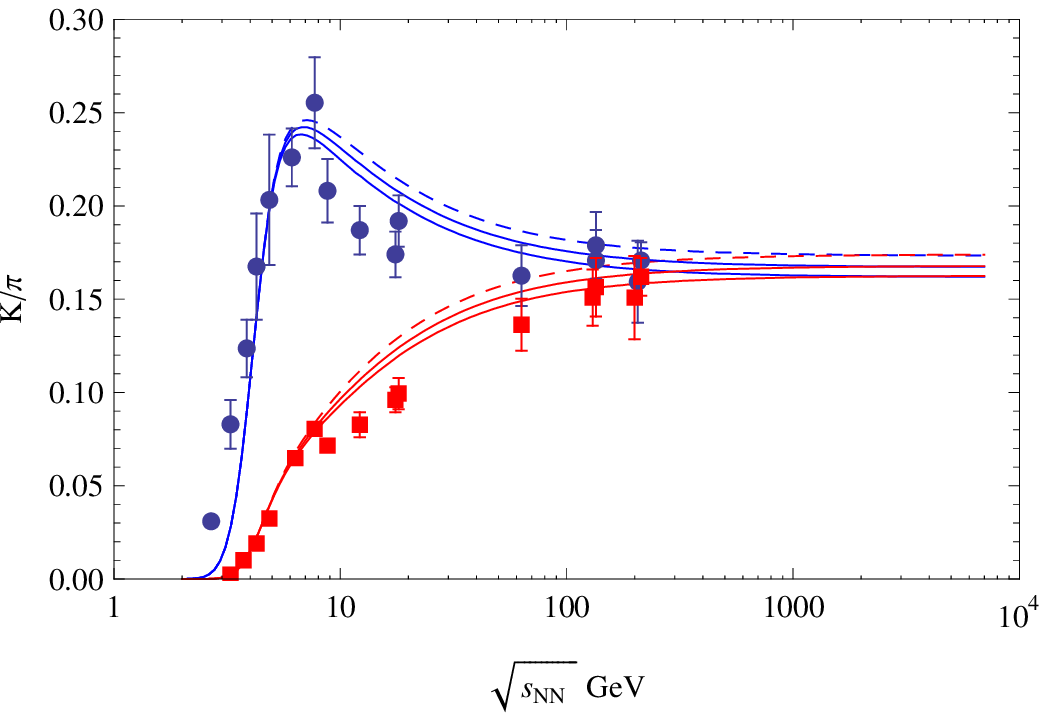}}
\end{center}
\caption{The $K/\pi$ ratio obtained in our models using as input
  the energy dependence of the freezeout values of $T$ and $\mu$ obtained
  in \cite{andronic} by fitting to the data exhibited. The first figure
  is for decay model 1 and the second for decay model 2. Only the results
  for $\Lambda=2000$ MeV are shown for clarity.}
\label{fg.kbypi}
\end{figure}


In HRG2 the yield computation is not so straightforward, since one
has to start with a model ${\cal N}_{ji}$ for all $H_i$. In this first
study we restrict ourselves to models where $H_j=\pi$, K or a generic baryon
(\ie, we lump the ground-state octet and decuplet baryons into one).
As a result, we cannot answer detailed questions about the yield,
but only questions of total particle multiplicity and $K/\pi$ or
meson/baryon ratios.  In general, we expect ${\cal N}_{ji}$ (for fixed
$H_j$) to depend on the mass of $H_i$, so, a generic model for decays
is 
\beq
   {\cal N}_{ji}=a^j_0+a^j_1 m_i+a^j_2 m_i^2+\cdots.
\label{decaymodel}\eeq
While ${\cal N}_{ji}$ may well depend on other quantum number of $H_i$,
we are neither able to confirm or rule this out. Therefore, we decided
to work with the simple model above in this paper.  The decays of the
heaviest hadrons are not yet studied well enough to provide further
constraints.  Further, incomplete data may contain unknown biases. Hence
we neglected the heaviest hadrons and constrained the model with data
for hadrons with masses up to 2000 MeV.

Due to the high threshold for the production of a baryon-antibaryon
pair, the decays of such particles do not involve such a pair in the
final state. Such pairs are rare also in the known decays of hadrons
with mass greater than 2000 MeV, being seen mainly in the final state
of strange mesons. The branching ratios to such pairs are still mostly
ill-measured. As a result, we are unable to make any testable model
about the expected occurrence of baryon-antibaryon pairs in decays of
the Hagedorn spectrum of resonances. We therefore take the simplest
possible model:
\beq
   {\cal N}_{Bi} = B_i,
\label{decaybaryon}\eeq
that the number of baryons in the final state of the decay of any particle
is equal to the baryon number of the resonance, $B_i$. When better data on
resonances between 2000 and 3000 GeV becomes available, the model can be
refined by adding higher terms from eq.\ (\ref{decaymodel}).

The production of $K\overline K$ pairs is also strongly suppressed in
the decays of hadrons; this is a statement of the OZI rule. Cases which
contradict this rule are known, but our attempts to make statistical
models of such decays falls on the rock of insufficient data. Thus, we
are forced to the rule that 
\beq
   {\cal N}_{Ki} = S_i,
\label{decaykaon}\eeq
the number of kaons produced in the decay of $H_i$ is equal to its
strangeness, $S_i$. We also examined a variant of this rule, which is
that the strangeness in the baryon sector percolates down to strange
baryons, and does not go into production of kaons. This corresponds
to the variant model ${\cal N}_{Ki}=S_i (1-\delta_{B_i,1})$. We refer
to this as the decay model 2, to distinguish it from the model in
eq.\ (\ref{decaykaon}), which we call decay model 1.

Finally, we consider the expected number of pions produced in the decays
of resonance. We fit the linear model
\beq
   {\cal N}_{\pi i} = b + a {\cal M}_i,
\label{decaypion}\eeq
where ${\cal M}_i$ is that part of the mass of $H_i$ which is available
for decay to pions. For unflavoured mesons, the available mass is the mass
of the particle, for strange mesons, the available mass is the difference
between the mass of a kaon and the mass of $H_i$, and for a baryon it
is the difference between the mass of $H_i$ and a typical baryon mass,
which we take to be 1000 MeV. This simple model is fitted to four sets of
data--- separately for strange and non-strange mesons and baryons. In
all these cases we found $b$ consistent with zero and $a\simeq1.5$
GeV$^{-1}$, and consistent with each other. Adding in a term quadratic
in masses to the model does not improve the fit significantly, so we
keep to the linear model above. It may be useful to note the following
implication: out of every GeV of rest mass of the higher resonances,
about 500 MeV is available as the kinetic energy of each pion.

This completes the specification of a simple model for the ${\cal
N}_{ji}$ in HRG2. Many more refinements and elaborations of the model
are possible, however, this model is sufficient for the computations
that we exhibit next. We extend the yield formula of eq.\ (\ref{yield})
to
\beq
   N_j(V,T,\mu) = \sum_i {\cal N}_{ji} N^{\rm th}_i(V,T,{\mu}),
\label{yield2}\eeq
where we use the empirically determined ${\cal N}_{ij}$ for $m_i<\Lambda$
and the model for $m_i>\lambda$. There are a very small number of baryons
whose masses are sufficiently well-known to be included in HRG1, but whose
decays are not very well known. For these we used the decay models.

The $K^\pm/\pi^\pm$ ratios have been determined at AGS \cite{ags}, SPS
\cite{sps} and RHIC \cite{rhic} energies. In order to compare with this
data we need to take into account the fact that the initial state has
no strangeness.  Due to strangeness conservation in strong interactions
this condition of zero overall strangeness has to be enforced as a
canonical constraint, \ie, the $K/\pi$ ratios have to be determined in
the canonical ensemble \cite{suhonen2,pbmrev}. We implement this in our
computations of hadron yields.

Our results for the $K/\pi$ ratio as a function of the beam energy
$\sqrt S$ are shown in Figure \ref{fg.kbypi}. In this computation we have
used the freezeout parameters deduced in \cite{andronic}. As expected,
the ratios of yields are a little higher in decay model 1. Note also
that there is a little difference between the predictions of HRG1 and
HRG2. For the same freezeout conditions, HRG2 gives a slightly smaller
value of the yield ratio. In principle, the freezeout conditions that
one deduces from data could be dependent on specifics of the hadron
gas model one uses.  Our results indicate that this dependence is at
best mild. To check this we have implemented the freezeout criterion of
\cite{freezeout}. Using this we find that the freezeout parameters change
only marginally from HRG1 to HRG2, consistent both with the conclusions
of \cite{freezeoutstable} and the results in Figure \ref{fg.kbypi}. In
future we plan a more detailed study of hadron yields, and the effects
of the Hagedorn spectrum on the chemical composition at freezeout in
baryon rich matter.

We conclude with a summary of our investigation into hadron resonance
gas models of the kind which have been used to explain hadron yields in
heavy-ion collisions. The thermodynamics of such resonance gases (HRG1) is
strongly dependent on the cutoff in the spectrum at temperatures of over
100 MeV. In this paper we have investigated an extension of the hadron
resonance gas models in which as yet unobserved resonances are included
using the Hagedorn model of the hadron spectrum (HRG2) with two different
densities of state (eqs.\ \ref{hg2} and \ref{hg2p}). We found that the
Hagedorn temperature, $\th$, is strongly dependent on details of the model
for the hadron spectrum. This implies that the present knowledge of the
spectrum of QCD is as yet unable to constrain string models of hadrons.

Each of the HRG2 models stabilizes the uncertainty in thermodynamics
below $T_c$ due to the cutoff in HRG1. However, the predictions of the
two variants of the model are different. While the model is defined up to
$\th$, we found that HRG2 gives unrealistic results for thermodynamics,
especially for $C_{BS}$, from $T_c$ to $\th$. Since there is no phase
transition in QCD, a failure of a model near and above $T_c$ also
implies a failure near and below $T_c$. This leads us to believe that
the accuracy of resonance gas models cannot be pushed much closer to
$T_c$ than the freezeout temperature. Of course, this leaves open the
interesting possibility that resonance gas models describe the physics
of transport that leads to freezeout, for example, the thickness of the
freezeout layer.

We also investigated observable quantities in heavy-ion collisions,
such as hadron yields.  In order to do that, we had to develop a
novel model for the decay of Hagedorn resonances.  We found that the
significant uncertainties that we saw in thermodynamics lead to rather
small changes in observables such as the $K/\pi$ yield ratio. This has
positive implications on efforts to explain hadron yields from heavy-ion
observation.

\end{document}